\documentclass[12pt]{report}
\usepackage{graphics}
\usepackage{float}
\usepackage{epsfig}
\usepackage{makeidx}
\usepackage{amssymb}
\usepackage{epsfig}
\usepackage{amsmath}
\usepackage{bm}
\usepackage{graphicx}
\usepackage{epstopdf}

\begin{document}

\vskip 0.5 in
\hskip -.2in{\Large{\bf{Effects of  Planar Periodic Stratified Chiral Nihility }}}\\
 \hskip -.2in{\Large{\bf{Structures   on Reflected and Transmitted Powers}}}\\
\vskip 0.2 in
\hskip -.2in{{\bf{Nayyar Abbas Shah$^1$, Faiz Ahmad$^2$, Aqeel A. Syed$^3$, Qaisar A. Naqvi$^4$}}}\\

\vskip 0.2 in \hskip -.2in{{{$^{1}$nayyarabbas07@yahoo.com, $^{2}$faizsolangi@gmail.com,}}}\\
\hskip -.2in{{{ $^{3}$aqeel@qau.edu.pk, $^{4}$qaisar@qau.edu.pk}}}\\

 \hskip -.2in{{{$^{1,3,4}$Department of Electronics, Quaid-i-Azam
University, Islamabad
45320, Pakistan}}}\\
\hskip -.2in{{{$^{2}$Department of Physics, CIIT,
Islamabad, Pakistan}}}\\

\vskip 0.2 in \centerline {\Large{\bf {Abstract}}} Behavior of
planar multilayer periodic structures due to plane wave excitation
has been studied using the transfer matrix method. Multilayer
structure is taken with periodicity two. That is,  layers at even
and odd locations repeat themselves. Layers at odd locations are of
chiral nihility metamaterial whereas three different cases  for
layers at even locations are considered, i.e., dielectric, chiral
and chiral nihility. Effects of polarization rotation due to the
optical activity is studied with respect to the angle of incidence
and frequency in terahertz domain. Chiral nihility introduces
property of transparency to the structure for normal incidence while
complete rejection is observed for chiral nihility-chiral nihility structure at oblique incidence .\\
\section{Introduction}
Objects in our universe can be classified in two groups: achiral and
chiral.  An object not superimposable  to its mirror image by any
translation or rotation is called chiral object. Achiral object are
non-chiral and can be superposed to its mirror image [1, 2]. A
metamaterial microscopically composed of chiral objects is called
chiral metamaterials. The constitutive relations for isotopic,
homogenous and non-diffusive chiral material are written below [3]:
\begin{eqnarray}
{\bf D}&=& \epsilon {\bf E}+(\chi-j\kappa)\sqrt{\epsilon_0\mu_0} {\bf H}\\
{\bf B}&=& \mu{\bf H}+(\chi+j\kappa)\sqrt{\epsilon_0\mu_0} {\bf E}
\end{eqnarray}
where $\kappa$ and $\chi$  are chirality and non-reciprocity
parameter respectively. Moreover
$\epsilon$ and $\mu$ are permittivity and permeability representing the electric and magnetic polarizability of the material.\\

\indent Lakhtakia proposed the concept of nihility metamaterial and
nihility means real parts of permittivity and permeability are zero
[4]. He studied the scattering of electromagnetic plane waves from a
cylinder composed of nihility metamaterial [5]. Later, Tretyakov et
al. extended the concept of nihility for the isotropic chiral
metamaterials [6]. Chiral nihility (CN) is a special case of chiral
metamaterial for which real parts of permittivity and permeability
are simultaneously zero at certain frequency, i.e.,
$\epsilon\longrightarrow0$ and $\mu\longrightarrow0$. The
constitutive relations for CN metamaterials are [6-8],
\begin{eqnarray}
{\bf D}&=& -j\kappa\sqrt{\epsilon_0\mu_0}{\bf H} \\
{\bf B}&=& j\kappa\sqrt{\epsilon_0\mu_0}{\bf E}.
\end{eqnarray}

Several researchers have studied the behavior of metamaterial
interfaces and structures when these are exposed to electromagnetic
excitation. Some special arrangements using metamaterials had been
proposed to fabricate absorbers, antenna radomes and cloaks [9-11].
Interaction of electromagnetic waves with chiral slabs with planner
interfaces had been studied for different applications [12-14].
 Rejection and tunneling of electromagnetic
fields from chiral and/or CN planar interfaces and waveguide
composed of planar CN metamaterials had also been studied [15, 16].
Reflection and transmission from CN slabs embedded in other
materials or backed by fractional dual interfaces are available in
[17, 18]. Khalid et al. had studied scattering of electromagnetic
waves from cylindrical DB boundaries in the presence of chiral and
CN metamaterials [19]. Taj et al. had studied the chiral and CN
metamaterials interfaces as focusing surfaces by using Maslov's
method [20]. Sabah et al. had studied the behavior of multilayered
structure. Sabah along with co-workers had calculated the response
of the chiral mirror by using transfer matrix method (TMM) [21].
Polarization rotator by using multilayered structures had been
studied by using conventional chiral and chiral metamaterial for
normal incidence [22]. Planner interfaces composed of chiral
metamaterials had studied for oblique incidence as filters in [23,
25].

In current paper, planar periodic multilayered structures having odd
number of slabs are analyzed. Two layers form a period and it is
assumed that layers at odd locations are of CN metamaterial. Three
different cases are considered for layers at even locations, i.e.,
dielectric, chiral, and CN.  Purpose of the study is to find the
effects of CN metamaterial on reflected and transmitted powers.

\section{Formulation} Consider planar
periodic stratified structure  of chiral metamaterials as shown in
Figure~1. Each slab is of infinite extent in x and y directions of
the Cartesian coordinate system. Slabs at odd locations have
identical constitutive parameters while slabs at even locations have
same values of constitutive parameters. Therefore, each slab at odd
location is labeled as A whereas each slab at even location is
labeled as B. It is also assumed that slab labeled A have higher
value of refractive index than slab labeled B. Constitutive
parameters for slab A are denoted by $(\epsilon_H, \mu_H, \kappa_H)$
and constitutive parameters for slab B are denoted by $(\epsilon_L,
\mu_L, \kappa_L)$.
 The stratified structure is placed in air
medium having constitutive parameters $(\epsilon_0, \mu_0)$.
\begin{figure}[H]
\centerline {\epsfxsize 4 in \epsfysize 2.5 in \epsfbox{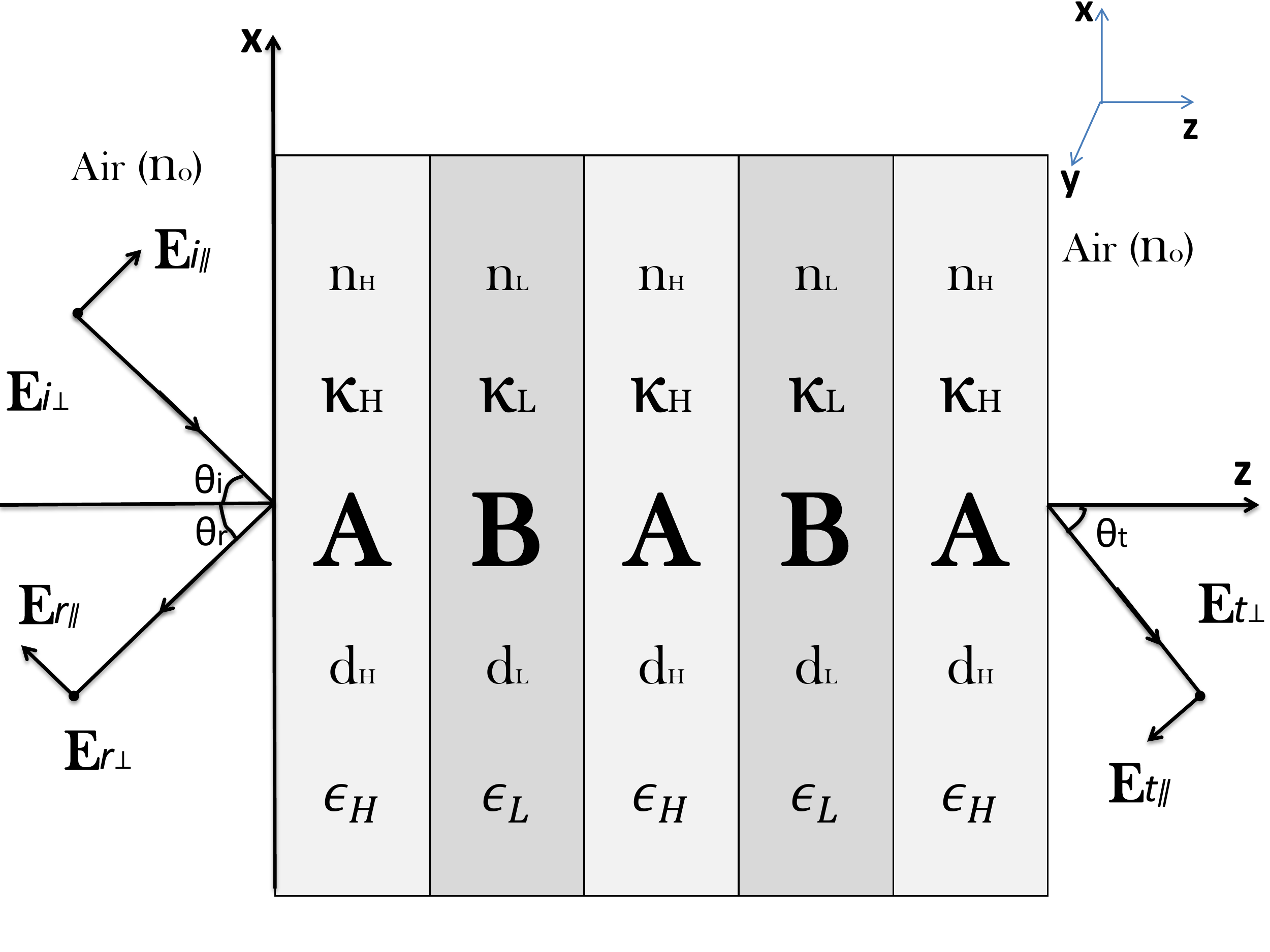}}
  \caption{Stratified metamaterials at oblique incidence}\label{$2.1$}
\end{figure}
\indent The structure is excited by a parallel polarized
monochromatic  plane wave. The expression for incident electric
field is given below:
\begin{eqnarray}
  {\bf E}_i &=& [E_{i\parallel}(\hat x \cos\theta_i+ \hat z \sin\theta_i)]\exp[-jk_{0}(z \cos\theta_i-x\sin\theta_i)]
\end{eqnarray}
where $k_0=\omega\sqrt{\epsilon_0\mu_0}$ and $\theta_i$ is the angle of incidence with respect to z-axis.\\
\indent  It is assumed that reflected and transmitted fields contain
both parallel and perpendicular components. Expressions for
reflected and transmitted waves, in terms of unknown coefficients,
are written as
\begin{eqnarray}
{\bf E}_r &=& [E_{r\parallel}( \hat x \cos\theta_r- \hat z \sin\theta_r)+E_{r\perp}\hat y]\exp[-jk_{0}(-z\cos\theta_r-x\sin\theta_r)]  \\
{\bf E_t} &=& [E_{t\parallel}( \hat x \cos\theta_t- \hat z \sin\theta_t)+ E_{t\perp}\hat y]\exp[-jk_{0}(z\cos\theta_t-x\sin\theta_t)]
\end{eqnarray}
$\theta_r$ and $\theta_t$ are angles of reflection and transmission in air with respect to z-axis.\\
\indent Electric and magnetic fields inside the $m$-th chiral slab
are written as linear combination of left circularly and right
circularly polarized plane waves. Both waves are propagating in
forward as well as backward directions, that is, in positive and
negative z-directions. Expressions for total electric field in
$m$-th chiral slab are given below [9]:
\begin{eqnarray*}
{\bf E_m^{+}}&=&{ {\bf E}_{mL}^+}\exp[-jk_{mL}(z\cos\theta_{mL}-x\sin\theta_{mL})]+{{\bf E}_{mR}^+}\exp[-jk_{mR}(z\cos\theta_{mR}-x\sin\theta_{mR})],  \\
{\bf E_m^{-}}&=&{ {\bf E}_{mL}^-}\exp[-jk_{mL}(-z\cos\theta_{mL}-x\sin\theta_{mL})]+{{\bf E}_{mR}^-}\exp[-jk_{mR}(-z\cos\theta_{mR}-x\sin\theta_{mR})],
\end{eqnarray*}
with\\
\begin{eqnarray}
{{\bf E}_{mL}^+} &=& {E_{mL}^+}({\hat x}\cos\theta_{mL}+{\hat z}\sin\theta_{mL}+j{\hat y}),\\
{{\bf E}_{mR}^+} &=& {E_{mR}^+}({\hat x}\cos\theta_{mR}+{\hat z}\sin\theta_{mR}-j{\hat y}),\\
{{\bf E}_{mL}^-} &=& {E_{mL}^-}({\hat z}\sin\theta_{mL}-{\hat x}\cos\theta_{mL}+j{\hat y}),\\
{{\bf E}_{mR}^-} &=& {E_{mR}^-}({\hat z}\sin\theta_{mR}-{\hat x}\cos\theta_{mR}-j{\hat y}).
\end{eqnarray}
Superscripts + and - are used to describe the waves propagating in
forward and backward directions, respectively. Moreover
$\theta_{mL}$ and $\theta_{mR}$ are used to describe the angles of
LCP and RCP waves with respect to the z-axis. The magnetic fields
outside and inside the chiral slab may be obtained from Maxwell curl
equations. $k_{mR}$ and $k_{mL}$ are wave numbers for LCP and RCP
waves, respectively and are given as,
\begin{eqnarray}
k_{mL} &=& \omega(-\kappa_i\sqrt{\epsilon_{0}\mu_{0}}+\sqrt{\epsilon_i\mu_i})\\
k_{mR} &=&
\omega(\kappa_i\sqrt{\epsilon_{0}\mu_{0}}+\sqrt{\epsilon_i\mu_i}),
\qquad i=H, L.
\end{eqnarray}
$E_{r\parallel}$, $E_{t\parallel}$, $E_{r\perp}$, $E_{t\perp}$,
${E_{mL}^\pm}$ and ${E_{mR}^\pm}$ are unknown coefficients to be
determined using the boundary conditions.
Transfer matrix method (TMM) is used to determine the reflected and
transmitted fields from the  planar stratified structure. First of
all, matching matrices are calculated by imposing the boundary
conditions at each interface. To relate fields at two consecutive
interfaces, propagation matrices are also computed. Product of these
matrices finally relates the reflected and transmitted fields with
the incident field in form of matrix equation as given below [25],
\begin{eqnarray}
 \left(
   \begin{array}{c}
     E_{i\parallel} \\
     E_{i\perp}\\
     E_{r\parallel} \\
     E_{r\perp}\\
   \end{array}
 \right),
 &=&
T
 \left(
   \begin{array}{c}
     E_{t\parallel} \\
     E_{t\perp}\\
   \end{array}
 \right)
\end{eqnarray}
where relating transfer matrix is given below,
\begin{eqnarray}
T &=& M_1.P_A.T_1^{m}.M_2 \\
T_1 &=& M_{AB}.P_B.M_{BA}.P_{A}
\end{eqnarray}
with $M_1$ and $M_2$ matching matrices which relate fields across
first and last interface of planar stratified structure,
respectively. In general, matching matrices $M_{ij}$ correspond to
an interface with slab $i$ on left side and slab $j$ on
right side of the interface. Propagation matrices are represented by $P_A$ and $P_B$ corresponding to slab A and slab B, respectively.\\
\section{Results and Discussions} In this section,
numerical results for three different structures are obtained to
note the effects of CN metamaterials on transmitted and reflected
powers. For all cases, parallel polarized plane wave
$(E_{i\perp}=0)$ is considered for incidence and structure have odd
numbers of slabs. In all cases, each slab have optical width
$\lambda_0$/4 except for CN case when slabs have physical width
equal to $\lambda_0$/4,  where $\lambda_0$ corresponds to the first
harmonic frequency of $f_0=$1THz.\\
\vskip 0.1 in
\subsection{CN-dielectric Structure}
\indent  In the first case, slabs labeled as A are of CN
metamaterial whereas  slabs labeled as B are of dielectric medium.
It is  assumed that structure is composed of five slabs. Each CN
slab has physical width of $\lambda_0/4$ and each dielectric slab
have thickness $\lambda_0/4n_d$, where $n_d$ is refractive index of
the dielectric slab. Figure 2 describes behavior of reflected power
versus frequency for normal incidence. It has sinusoidal behavior
with maximum amplitude equal to 40 percent of the incident power.
Reflected power approaches to zero at 1.0, 2.0, 3.0 and 4.0THz
frequencies. Figure 3 shows the behavior of transmitted power versus
frequency. Transmitted power has both cross and co-polarized
components and these are shown by dotted and solid lines,
respectively. It is also observed that rotation is $45^o$ at 1.0 and
3.0THz, and complete rotation is noted at
2THz for normal incidence.\\
\indent In Figure 4 and Figure 5, reflected and transmitted powers
versus frequency for oblique incidence are plotted. It is noted that
total reflection is observed at higher frequencies but some ripples
are also noted in the
reflected and transmitted powers at lower values of frequencies.\\
\indent Figure 6 shows reflected power versus angle of incidence.
Total reflection is noted for  $\theta_i>22^o$ for fixed frequency
of 1THz. The increasing trend for co-polarized component is also
observed and it takes  value unity from $\theta_i=60^o$ to
$\theta_i=90^o$. Corresponding transmitted power,  shown in Figure
7, has both cross and co-polarized components up to $\theta_i=22^0$
having maximum rotation at $15^o$ and becomes zero for next values.
\subsection{CN-CN structure} For second case, it is assumed
that both slabs A and B are of CN metamaterials with same value of
chirality but having different limiting values of permittivity and
permeability. Physical width of each slab is equal to $\lambda_0/4$.
In Figure~8 and Figure 9, it is observed that total power is
transmitted through the structure but due to optical activity of the
chiral material,  the plane of polarization gets rotated. Rotation
is $90^0$ at 2THz frequency for the chirality parameter
$\kappa_H=\kappa_L$=0.1\\
\indent In Figure 10 and Figure 11, behavior of CN-CN structure is
studied versus frequency for oblique incidence. For these figures,
$\theta_i=45^0$ is considered and total reflection is noted after
fraction of 1THz and total transmission is
noted for 0.2THz.\\
\indent  Next two figures describe behavior of reflected and
transmitted power versus angle of incidence. In Figure 12, complete
reflected power is noted from $\theta_i=15^o$ to $\theta_i=90^o$
having maximum contribution of co-polarized component and cross
component has largest contribution at $\theta_i=10^o$. In Figure 13,
transmitted power is observed only
at lower values of angle of incidence with major contribution from co-polarized component.\\
\subsection{CN-chiral structure}
In third case, we consider that the slab at odd positions are of
chiral metamaterial. Chiral slabs have optical width equal to
quarter wavelength and lower chirality value than CN slabs. In
Figure 14 and Figure 15, it is noted that reflected power has small
percentage of incident power and transmitted power has major
contribution of the cross polarized component for small values of
angle of incidence. Whereas at higher degrees of incidence angle
reflected power starts increasing and transmitted power shows
decaying trend.
\begin{figure}[H]
 \centerline {\epsfxsize 4.5 in \epsfysize 2.8 in \epsfbox{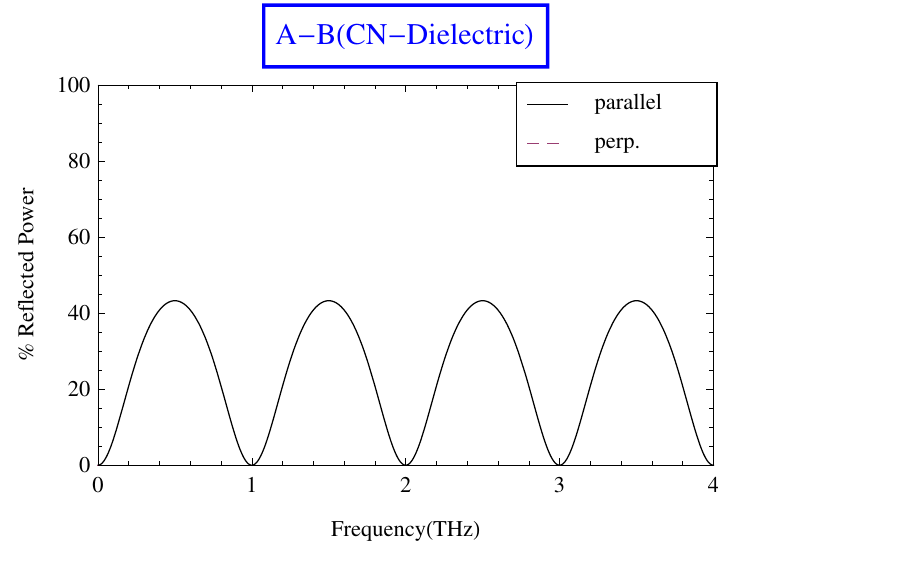}}
  \caption{$\epsilon_H=1.6*10^{-4}$, $\mu_H= 1*10^{-5}$, $n_L=2.2$, $d_H=|n_L|d_L=\lambda_0/4$, $\kappa_H=0.167$}\label{$$}
\end{figure}
\begin{figure}[H]
 \centerline {\epsfxsize 4.5 in \epsfysize 2.8 in \epsfbox{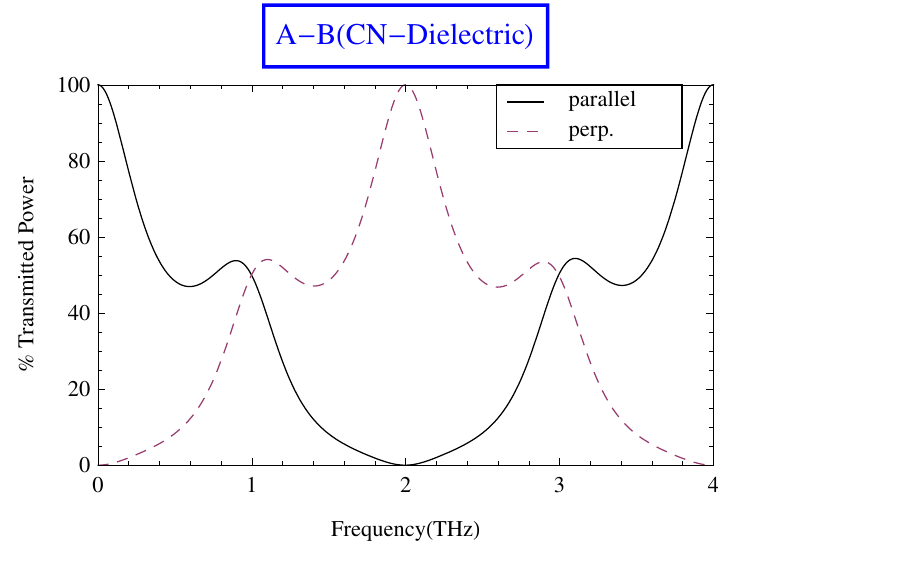}}
  \caption{$\epsilon_H=1.6*10^{-4}$, $\mu_H= 1*10^{-5}$, $n_L=2.2$, $d_H=|n_L|d_L=\lambda_0/4$, $\kappa_H=0.167$}\label{$$}
\end{figure}
\begin{figure}[H]
 \centerline {\epsfxsize 4.5 in \epsfysize 2.8 in \epsfbox{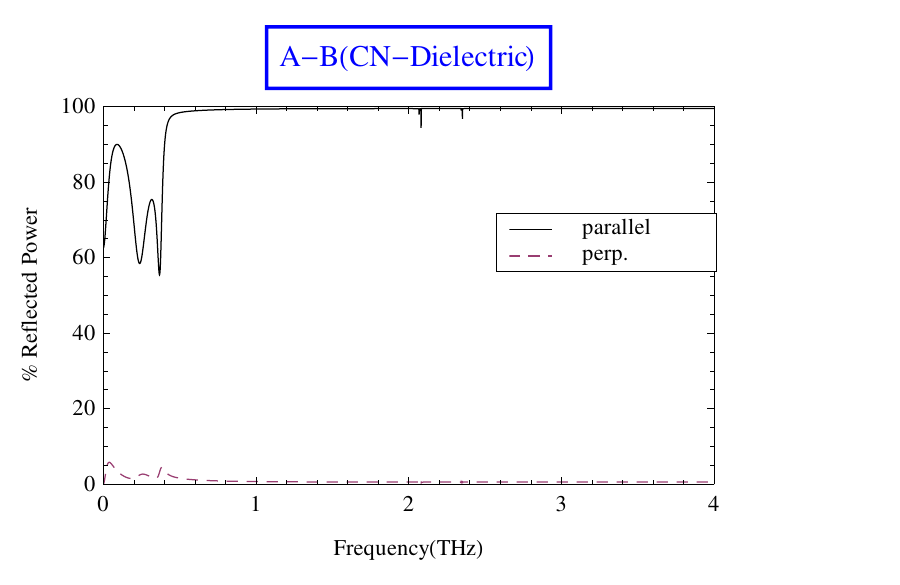}}
  \caption{$\epsilon_H=1.6\times10^{-4}$, $\mu_H=1\times10^{-5}$, $n_L=2.2$, $d_H=|n_L|d_L=\lambda_0/4$,
  $\kappa_H=\kappa_L=\kappa=0.1$, $\theta_i=70^0$}\label{$2.1$}
\end{figure}
\begin{figure}[H]
 \centerline {\epsfxsize 4.5 in \epsfysize 2.8 in \epsfbox{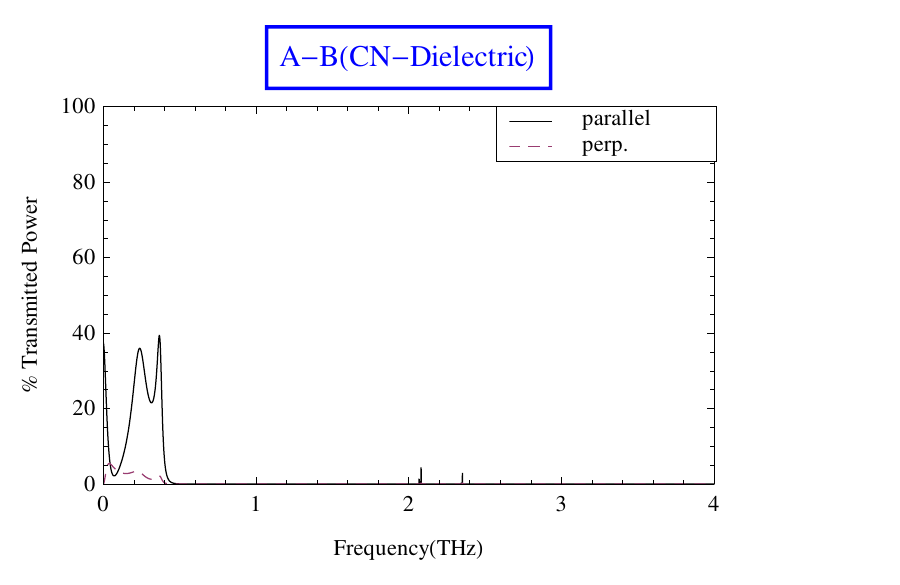}}
  \caption{$\epsilon_H=1.6\times10^{-4}$, $\mu_H=1\times10^{-5}$, $n_L=2.2$, $d_H=|n_L|d_L=\lambda_0/4$,
  $\kappa_H=\kappa_L=\kappa=0.1$, $\theta_i=70^0$}\label{$2.1$}
\end{figure}
\begin{figure}[H]
 \centerline {\epsfxsize 4.5 in \epsfysize 2.8 in \epsfbox{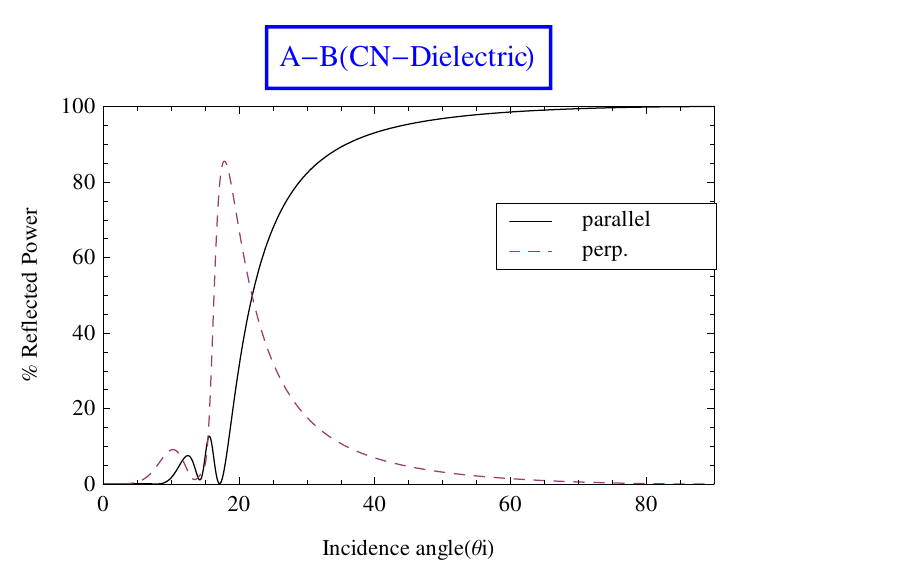}}
  \caption{$\epsilon_H=1.6\times10^{-4}$, $\mu_H=1\times10^{-5}$, $n_L=2.2$, $d_H=|n_L|d_L=\lambda_0/4$,
  $\kappa_H=0.1$, $f/f_0=1$}\label{$2.1$}
\end{figure}
\begin{figure}[H]
 \centerline {\epsfxsize 4.5 in \epsfysize 2.8 in \epsfbox{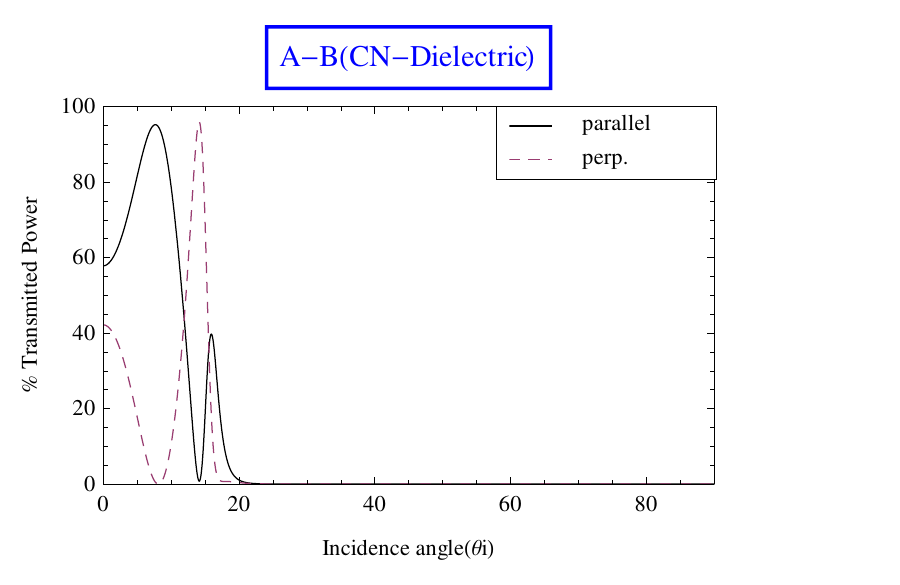}}
  \caption{$\epsilon_H=1.6\times10^{-4}$, $\mu_H=1\times10^{-5}$, $n_L=2.2$, $d_H=|n_L|d_L=\lambda_0/4$, $\kappa_H=0.1$, $f/f_0=1$}\label{$2.1$}
\end{figure}
\begin{figure}[H]
 \centerline {\epsfxsize 4.5 in \epsfysize 2.8 in \epsfbox{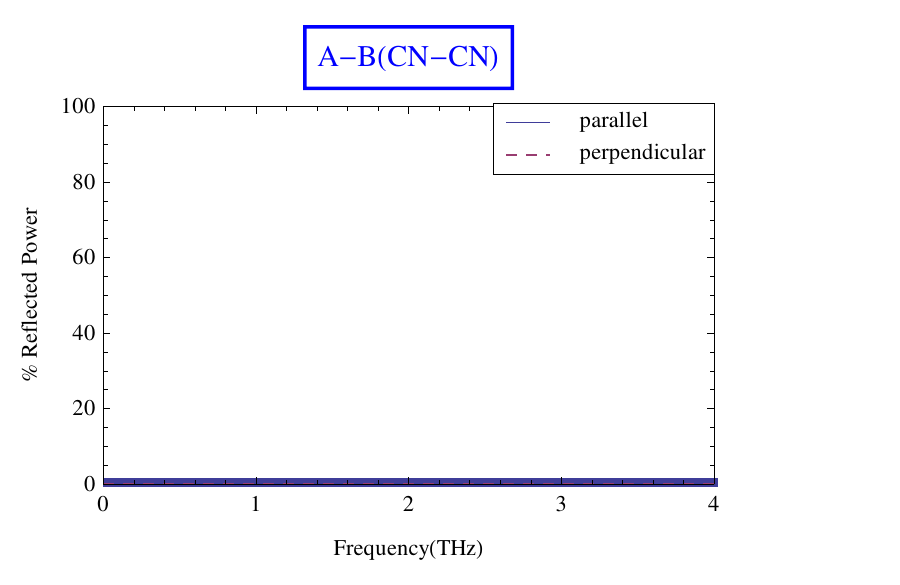}}
  \caption{$\epsilon_H=1.6\times10^{-4}$, $\epsilon_L=2.5\times10^{-5}$, $d_H=d_L=\lambda_0/4$, $\kappa_H=\kappa_L=0.1$}\label{$*$}
\end{figure}
 \begin{figure}[H]
 \centerline {\epsfxsize 4.5 in \epsfysize 2.8 in \epsfbox{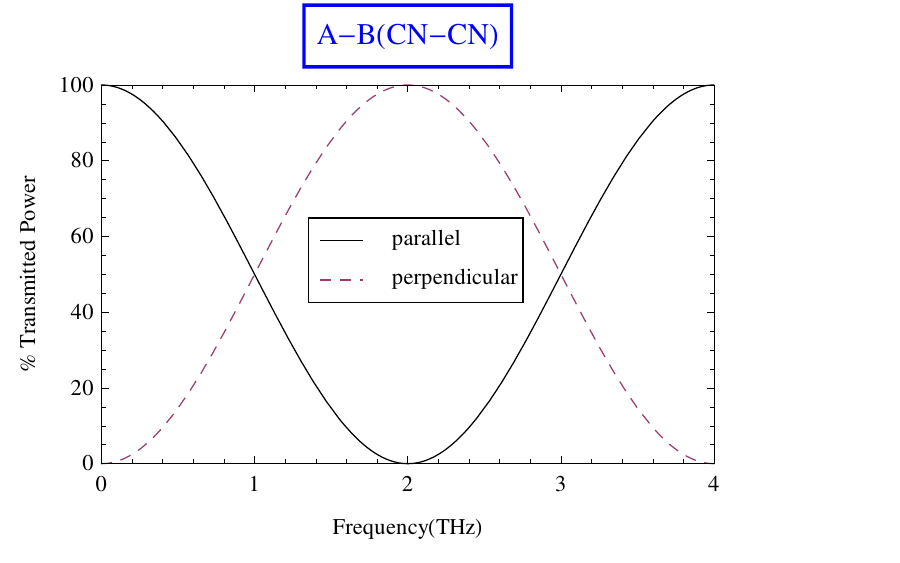}}
  \caption{$\epsilon_H=1.6\times10^{-4}$, $\epsilon_L=2.5\times10^{-5}$, $d_H=d_L=\lambda_0/4$, $\kappa_H=\kappa_L=0.1$}\label{$*$}
\end{figure}
\begin{figure}[H]
 \centerline {\epsfxsize 4.5 in \epsfysize 2.8 in \epsfbox{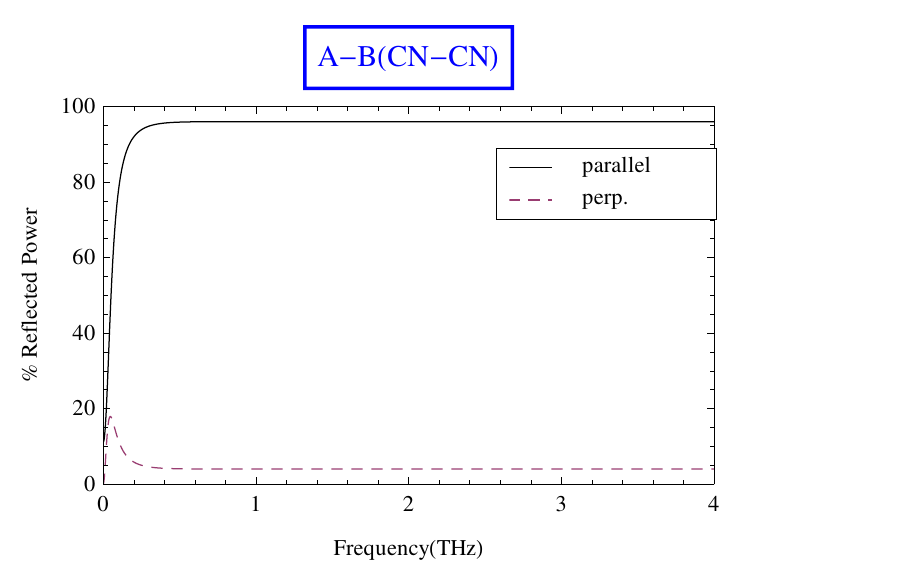}}
  \caption{$\epsilon_H=1.6\times10^{-4}$, $\epsilon_L=2.5\times10^{-5}$, $\mu_H=\mu_L=10^{-5}$, $d_H=d_L=\lambda_0/4$,
  $\kappa_H=\kappa_L=0.1$, $\theta_i=45$}\label{$2.1$}
\end{figure}
\begin{figure}[H]
 \centerline {\epsfxsize 4.5 in \epsfysize 2.8 in \epsfbox{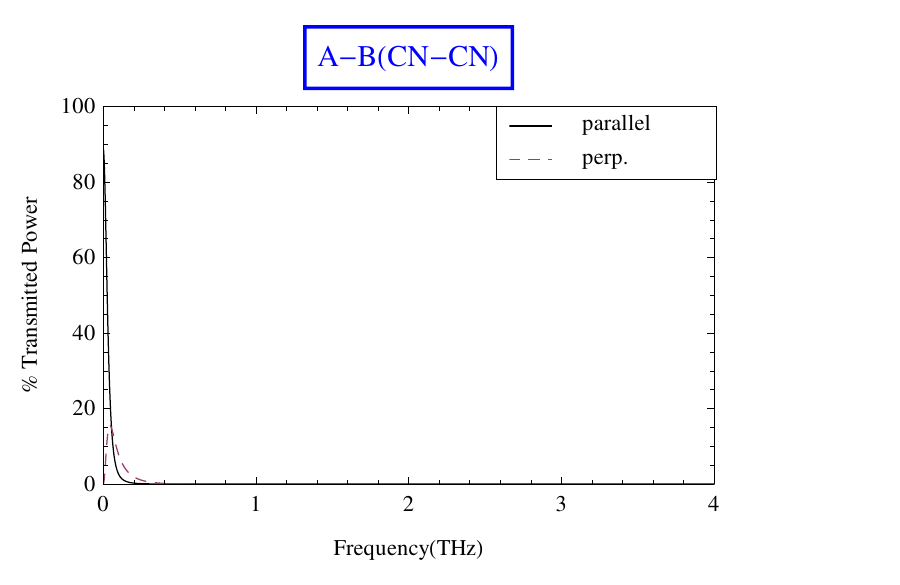}}
  \caption{$\epsilon_H=1.6\times10^{-4}$, $\epsilon_L=2.5\times10^{-5}$, $\mu_H=\mu_L=10^{-5}$, $d_H=d_L=\lambda_0/4$, $\kappa_H=\kappa_L=0.1$, $\theta_i=15$}\label{$2.1$}
\end{figure}
\begin{figure}[H]
 \centerline {\epsfxsize 4.5 in \epsfysize 2.8 in \epsfbox {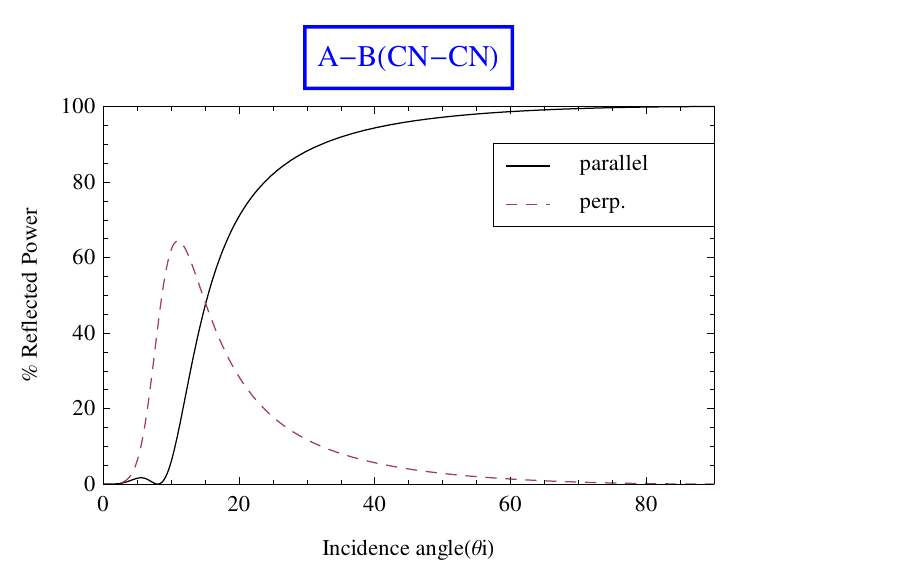}}
  \caption{$\epsilon_H=1.6\times10^{-4}$, $\epsilon_L=2.5\times10^{-5}$, $\mu_H=\mu_L=10^{-5}$, $d_H=d_L=\lambda_0/4$, $\kappa_H=\kappa_L=0.1$, $f/f_0=1$}\label{$2.1$}
\end{figure}
\begin{figure}[H]
 \centerline {\epsfxsize 4.5 in \epsfysize 2.8 in \epsfbox{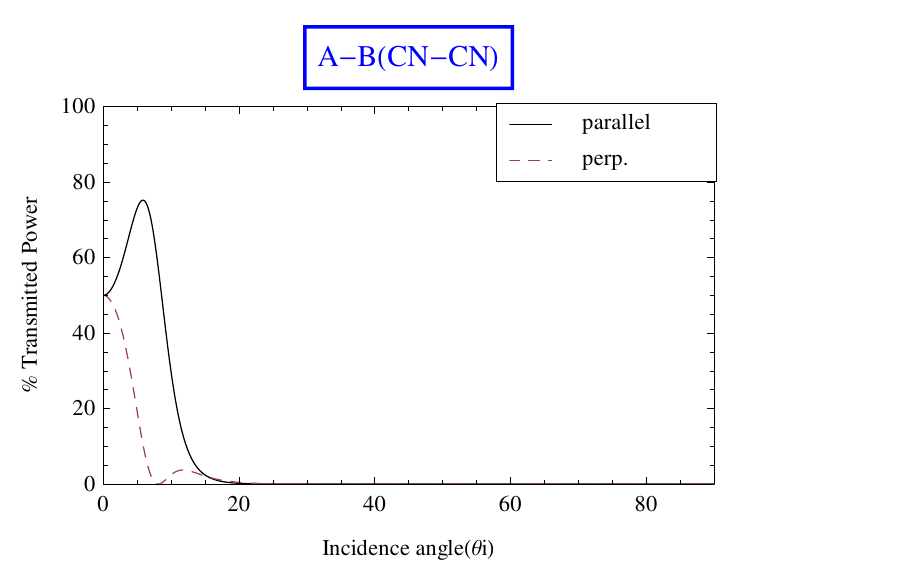}}
  \caption{$\epsilon_H=1.6\times10^{-4}$, $\epsilon_L=2.5\times10^{-5}$, $d_H=d_L=\lambda_0/4$, $\mu_H=\mu_L=10^{-5}$, $\kappa_H=\kappa_L=0.1$, $f/f_0=1$}\label{$2.1$}
\end{figure}
\begin{figure}[H]
 \centerline {\epsfxsize 4.5 in \epsfysize 2.8 in \epsfbox {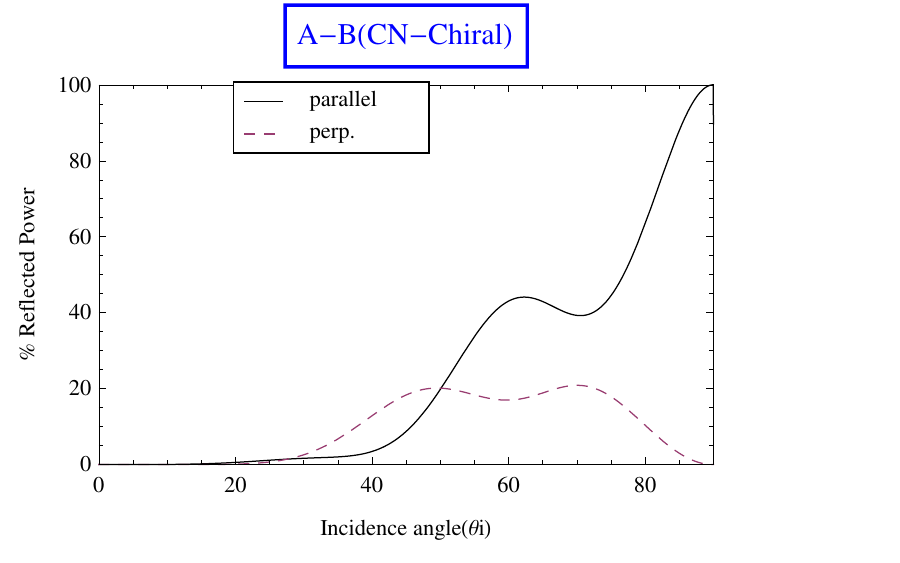}}
  \caption{$\epsilon_H=1.6\times10^{-4}$, $\epsilon_L=2.5\times10^{-5}$, $\mu_H=\mu_L=10^{-5}$, $d_H=d_L=\lambda_0/4$, $\kappa_H=\kappa_L=0.1$, $f/f_0=1$}\label{$2.1$}
\end{figure}
\begin{figure}[H]
 \centerline {\epsfxsize 4.5 in \epsfysize 2.8 in \epsfbox{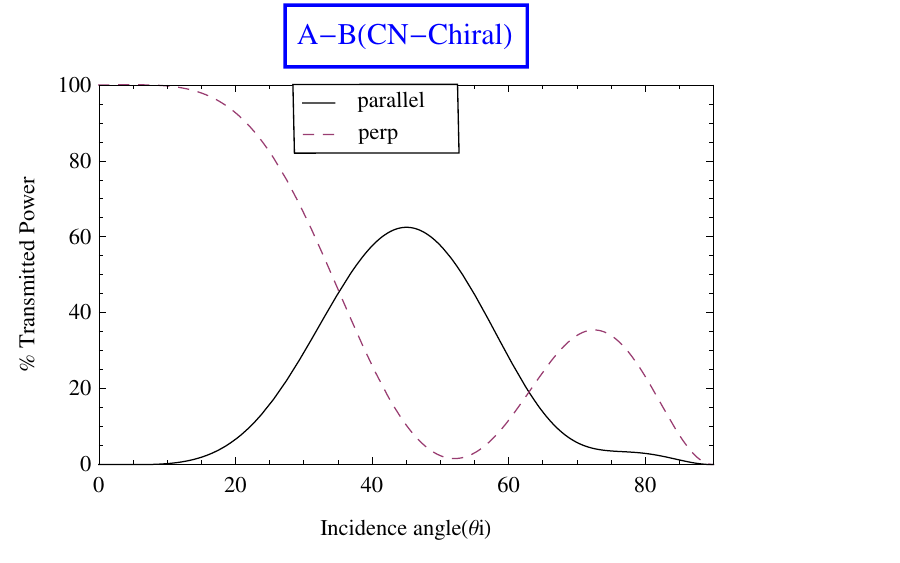}}
  \caption{$\epsilon_H=1.6\times10^{-4}$, $\epsilon_L=2.5\times10^{-5}$, $d_H=d_L=\lambda_0/4$, $\mu_H=\mu_L=10^{-5}$, $\kappa_H=\kappa_L=0.1$, $f/f_0=1$}\label{$2.1$}
\end{figure}
\section{Conclusions} \indent
Behavior of planar multilayer periodic chiral nihility structures
due to plane wave excitation has been studied using the transfer
matrix method. Each multilayer structure is taken with periodicity
two. First of all, behavior of CN-dielectric structure has been
studied for the case of normal incidence. Total transmission  as
well as maximum rotation of power happens at 2THz. This stratified
structure shows pronounced effects for oblique incidence. Total
reflection of power for higher frequencies and small ripples, both
in reflected and transmitted powers, for lower frequencies at
oblique incidence are noted. Reflected power approaches to unity by
increasing the value of angle of incidence and corresponding
transmitted power has major part for small values of incidence
angle.

Complete transparent behavior of CN-CN structure for all frequencies
has observed for normal incidence. Chirality introduces polarization
rotation having maximum rotation at even harmonics. But for oblique
incidence, total reflection from same structure has been observed.
At the end, behavior of CN-chiral structure has been studied and
total rejection is observed for small incidence angles having major
contribution of cross polarized components.

\end{document}